\documentclass[12pt]{article}

\usepackage{amsmath}
\usepackage{epsfig}

\begin{document}

\renewcommand{\thefootnote}{\fnsymbol{footnote}}

\begin{titlepage}

\begin{center}
\vspace*{.8 in} {\Large\bf Noise-Based Switches and Amplifiers
for}
\\ {\Large\bf Gene Expression}\\ \vspace*{0.6 in} {\large Jeff
Hasty$^{1}$, Joel Pradines$^{1}$, Milos Dolnik$^{1,2}$ and
J.J.~Collins$^{1}$}\\ \vspace*{0.6 in} {\large September 23, 1999}
\\ \vspace*{0.6 in} {$^{1}$Center for BioDynamics and Dept.~of
Biomedical Engineering, Boston University, 44 Cummington St.,
Boston, MA 02215, U.S.A.} \\ \vspace*{0.2 in} {$^{2}$Dept.~of
Chemistry and Center for Complex Systems, Brandeis University,
Waltham, MA 02454, U.S.A.}
\end{center}

\end{titlepage}

\newpage

\noindent {\bf ABSTRACT \hspace{0.1in} The regulation of cellular
function is often controlled at the level of gene transcription.
Such genetic regulation usually consists of interacting networks,
whereby gene products from a single network can act to control
their own expression or the production of protein in another
network. Engineered control of cellular function through the
design and manipulation of such networks lies within the
constraints of current technology. Here we develop a model
describing the regulation of gene expression, and elucidate the
effects of noise on the formulation. We consider a single network
derived from bacteriophage $\lambda$, and construct a
two-parameter deterministic model describing the temporal
evolution of the concentration of $\lambda$ repressor protein.
Bistability in the steady-state protein concentration arises
naturally, and we show how the bistable regime is enhanced with
the addition of the first operator site in the promotor region. We
then show how additive and multiplicative external noise can be
used to regulate expression. In the additive case, we demonstrate
the utility of such control through the construction of a protein
switch, whereby protein production is turned ``on'' and ``off''
using short noise pulses. In the multiplicative case, we show that
small deviations in the transcription rate can lead to large
fluctuations in the production of protein, and describe how these
fluctuations can be used to amplify protein production
significantly. These novel results suggest that an external noise
source could be used as a switch and/or amplifier for gene
expression. Such a development could have important implications
for gene therapy.}\normalsize

\vspace{0.5cm}

\baselineskip=24pt

\newpage

\section*{Introduction}
Regulated gene expression is the process through which cells
control fundamental functions, such as the production of enzymatic
and structural proteins, and the time sequence of this production
during development~\cite
{Dickson,Davidson1}. Many of these
regulatory processes take place at the level of gene transcription
\cite{GenesVI}, and there is evidence that the underlying
reactions governing transcription can be affected by external
influences from the environment~\cite{harada}.

As experimental techniques are increasingly capable of providing
reliable data pertaining to gene regulation, theoretical models
are becoming important in the understanding and manipulation of
such processes. The most common theoretical approach is to model
the interactions of elements in a regulatory network as
biochemical reactions. Given such a set of chemical reactions, the
individual jump processes (i.e., the creation or destruction of a
given reaction species) and their associated probabilities are
considered.  In its most general form, this often leads to a type
of Monte Carlo simulation of the interaction
probabilities~\cite{arkin}. Although this approach suffers from a
lack of analytic tractability, its strength is its completeness --
fluctuations in species' concentrations are embedded in the
modeling process. These internal fluctuations are important for
systems containing modest numbers of elements, or when the volume
is small.

Rate equations originate as a first approximation to such a
general approach, whereby internal fluctuations are ignored. These
deterministic differential equations describe the evolution of the
mean value of some property of the set of reactions, typically the
concentrations of the various elements involved.  The existence of
positive or negative feedback in a regulatory network is thought
to be common~\cite{shapiro}, and, within the reaction framework,
feedback leads to nonlinear rate equations~\cite{keller}.

Noise in the form of random fluctuations arises in these systems
in one of two ways. As discussed above, {\em internal} noise is
inherent in the biochemical reactions. Its magnitude is
proportional to the inverse of the system size, and its origin is
often thermal. On the other hand, {\em external} noise originates
in the random variation of one or more of the externally set
control parameters, such as the rate constants associated with a
given set of reactions. If the noise source is small, its effect
can often be incorporated {\em post hoc} into the rate equations.
In the case of internal noise, this is done in an attempt to
recapture the lost information embodied in the rate equation
approximation.  But in the case of external noise, one often
wishes to introduce some new phenomenon where the details of the
effect are not precisely known.  In either case, the governing
rate equations are augmented with additive or multiplicative
stochastic terms. These terms, viewed as a random perturbation to
the deterministic picture, can induce various effects, most
notably the switching between potential attractors (i.e., fixed
points, limit cycles, chaotic attractors)~\cite{horsthemke}.

While impressive progress has been made in genome sequencing and
the understanding of certain qualitative features of gene
expression, there have been comparatively few advancements in the
quantitative understanding of genetic networks.  This is due to
the inherent complexity of such biological systems. In this work,
we adopt an engineering approach in studying a solitary gene
network. We envision that a plasmid, or {\em genetic
applet}~\cite{collins}, containing a small, self-contained gene
regulatory network, can be designed and studied in isolation. Such
an approach has two distinct advantages. First, since the approach
is inherently reductionist, it can make gene network problems
tractable and thus more amenable to a mathematical formulation.
Secondly, such an approach could form the basis for new techniques
in the regulation of {\em in vivo} gene networks, whereby a
genetic applet is designed to control cellular function.

In this paper, we develop a model describing the dynamics of
protein concentration in such a genetic applet, and demonstrate
how external noise can be used to control the network. Although
our results are general for networks designed with positive
autoregulation, we ground the discussion by considering an applet
derived from the promotor region of bacteriophage $\lambda$. Since
the range of potentially interesting behavior is wide, we focus
primarily on the steady-state mean value of the concentration of
the $\lambda$ repressor protein. This choice is motivated by
experiment; detailed dynamical information is still rather
difficult to obtain, as are statistical data concerning higher
moments. We show how an additive noise term can be introduced to
our model, and how the subsequent Langevin equation is analyzed by
way of transforming to an equation describing the evolution of a
probability function. We then obtain the steady-state mean
repressor concentration by solving this equation in the long time
limit, and discuss its relationship to the magnitude of the
external perturbation.  This leads to a potentially useful
application, whereby one utilizes the noise to construct a genetic
switch. We then consider noise at the level of transcription,
where noise enters the formulation in a multiplicative manner. As
in the additive case, we transform to an equation describing a
probability distribution, and solve for the steady-state mean
concentration as a function of noise strength. Finally, we
demonstrate how such a noise source can be used to amplify the
repressor concentration by several orders of magnitude.

\section*{A Model for Repressor Expression}
In the context of the lysis-lysogeny pathway in the $\lambda$
virus, the autoregulation of $\lambda$ repressor expression is
well-characterized~\cite{ptashne}. In this section, we present two
models describing the regulation of such a network. We envision
that our system is a plasmid consisting of the $P_R-P_{RM}$
operator region and components necessary for transcription,
translation, and degradation.

Although the full promotor region in $\lambda$ phage contains the
three operator sites known as OR1, OR2, and OR3, we first consider
a mutant system whereby the operator site OR1 is absent from the
region. The basic dynamical properties of this network, along with
a categorization of the biochemical reactions, are as
follows~\cite{ptashne}. The gene {\it cI} expresses repressor
(CI), which in turn dimerizes and binds to the DNA as a
transcription factor. In the mutant system, this binding can take
place at one of the two binding sites OR2 or OR3. (Here, we ignore
nonspecific binding.) Binding at OR2 enhances transcription, which
takes place downstream of OR3, while binding at OR3 represses
transcription, effectively turning off production.

The chemical reactions describing the network are naturally
divided into two categories -- fast and slow. The fast reactions
have rate constants of order seconds, and are therefore assumed to
be in equilibrium with respect to the slow reactions, which are
described by rates of order minutes. If we let $X$, $X_2$, and $D$
denote the repressor, repressor dimer, and DNA promoter site,
respectively, then we may write the equilibrium reactions
\begin{eqnarray}
 2X & \overset{K_{1}}{{\rightleftharpoons}} & X_2
\label{fast} \\
 D+X_2 & \overset{K_{2}}{{\rightleftharpoons}} & DX_2
 \nonumber \\
 D+X_2 & \overset{K_{3}}{{\rightleftharpoons}} &
 DX_2^*
 \nonumber \\
 DX_2 + X_2 & \overset{K_{4}}{{\rightleftharpoons}}
 & DX_2X_2 \nonumber
\end{eqnarray}
where the $DX_2$ and $DX_2^*$ complexes denote binding to the OR2
or OR3 sites, respectively, $DX_2X_2$ denotes binding to both
sites, and the $K_i$ are forward equilibrium constants. We let
$K_3=\sigma_1K_2$ and $K_4=\sigma_2K_2$, so that $\sigma_1$ and
$\sigma_2$ represent binding strengths relative to the dimer-OR2
strength.

The slow reactions are transcription and degradation,
\begin{eqnarray}
DX_2 + P & \overset{k_{t}}{\rightarrow} & DX_2 + P + nX
\label{slow}
\\
 X & \overset{k_{d}}{\rightarrow} & A \nonumber
\end{eqnarray}
where $P$ denotes the concentration of RNA polymerase and $n$ is
the number of proteins per mRNA transcript.  These reactions are
considered irreversible.

If we consider an {\em in vitro} system with high copy-number
plasmids\footnote{This assumption is necessary since the number of
relevant molecules per cell is small {\em in vivo}.  Since there
are many cells, we could alternatively use state probabilities as
dynamical variables describing an {\em in vivo} system.}, we may
define concentrations as our dynamical variables. Letting $x=[X]$,
$y=[X_2]$, $d=[D]$, $u=[DX_2]$, $v=[DX_2^*]$, and $z=[DX_2X_2]$,
we can write a rate equation describing the evolution of the
concentration of repressor,
\begin{equation}
 \dot x =  -2k_{1}x^2 + 2k_{-1}y + nk_{t}p_{0}u - k_{d}x + r
  \label{xeqn}
\end{equation}
where we assume that the concentration of RNA polymerase $p_0$
remains constant during time.  The parameter $r$ is the basal rate
of production of CI, i.e., the expression rate of the {\it cI}
gene in the absence of a transcription factor.

We next eliminate $y$, $u$, and $d$ from Eq.~(\ref{xeqn}) as
follows. We utilize the fact that the reactions in
Eq.~(\ref{fast}) are fast compared to expression and degradation,
and write algebraic expressions
\begin{eqnarray}
 y & = & K_1x^2 \label{approx1} \\
 u & = & K_{2}dy =  K_1K_2dx^2 \nonumber \\
 v & = & \sigma_1K_{2}dy =  \sigma_1K_1K_2dx^2 \nonumber \\
 z & = & \sigma_2K_{2}uy = \sigma_2(K_1K_2)^2dx^4 \nonumber
\end{eqnarray}
Further, the total concentration of DNA promoter sites $d_{T}$ is
constant, so that
\begin{equation}
d_T=d+u+v+z=d(1+(1+\sigma_1)K_1K_2x^2+\sigma_2K_1^2K_2^2x^4)
\end{equation}
Under these assumptions, Eq.~(\ref{xeqn}) becomes
\begin{equation}
\dot{x} =
\frac{nk_tp_0d_tK_1K_2~x^2}{1+(1+\sigma_1)K_1K_2x^2+\sigma_2K_1^2K_2^2x^4}
- k_d x + r
\end{equation}
Without loss of generality, we may eliminate two of the parameters
in Eq.~(\ref{xeqn}) by rescaling the repressor concentration $x$
and time. To this end, we define the dimensionless variables
$\widetilde{x}=x \sqrt{K_1K_2}$ and $\widetilde{t}=t
(r\sqrt{K_1K_2})$. Upon substitution into Eq.~(\ref{xeqn}), we
obtain
\begin{equation}
\dot x = \frac {\alpha~x^2}{1+(1+\sigma_1)x^2+\sigma_2x^4} -
\gamma~x + 1 \label{nodimx}
\end{equation}
where the time derivative is with respect to $\widetilde{t}$ and
we have suppressed the overbar on $x$. The dimensionless parameter
$\alpha \equiv n k_t p_0 d_T/ r$ is effectively a measure of the
degree in which the transcription rate is increased above the
basal rate by repressor binding, and $\gamma \equiv k_d /
(r\sqrt{K_1K_2})$ is proportional to the relative strengths of the
degradation and basal rates.

For the mutant operator region of $\lambda$ phage, we have
$\sigma_1 \sim 1$ and $\sigma_2 \sim 5$~\cite{ptashne,johnson}, so
that the two parameters $\alpha$ and $\gamma$ in
Eq.~(\ref{nodimx}) determine the steady-state concentration of
repressor. For this equation, there are two types of behavior. For
one set of parameter values, we have monostability, whereby all
initial concentrations evolve to the same fixed-point value.  For
another set, we have three fixed points, and the initial
concentration will determine which steady state is selected.
Additionally, in the multiple fixed-point regime, stability
analysis indicates that the middle fixed point $x_m$ is unstable,
so that all initial values $x<x_m$ will evolve to the lower fixed
point, while those satisfying $x>x_m$ will evolve to the upper.
This bistability arises as a consequence of the competition
between the production of $x$ along with dimerization and its
degradation. For certain parameter values, the initial
concentration is irrelevant, but for those that more closely
balance production and loss, the final concentration is determined
by the initial value.

Graphically, we can see how bistability arises in
Eq.~(\ref{nodimx}) by setting $\alpha~x^2 /(1+2~x^2+5~x^4)=
\gamma~x -1$.  In Fig.~1A we plot the functions $\alpha~x^2 /
(1+2~x^2+5~x^4)$ and $\gamma~x -1$ for fixed $\alpha$ and several
values of the slope $\gamma$. We see that for $\gamma$ small
(whereby degradation is minimal compared with production), there
is one possible steady-state value of $x$~(and therefore CI). As
we increase $\gamma$ above some critical value $\gamma_L$, we
observe that three fixed-point values appear. As we increase
$\gamma$ still further beyond a second critical value $\gamma_U$,
the concentration ``jumps'' to a lower value and the system
returns to a state of monostability.

The preceding ideas lead to a plausible method whereby the system
may be experimentally probed for bistability.  We envision that
$\alpha$ is fixed by the transcription rate and DNA binding site
concentration, and that the degradation parameter $\gamma$ is an
adjustable control. Beginning with a low initial value of
$\gamma=\gamma_0=5$, we slowly increase the degradation rate. The
effect is illustrated in Fig.~1B. We see that as $\gamma$ is
slowly increased, the concentration of CI slowly decreases as the
system tracks the fixed point.  Then, at the moment when $\gamma$
is greater than $\gamma_U$, the concentration abruptly jumps to a
lower value, followed by a further slow increase. Now suppose we
reverse course, and begin to decrease $\gamma$. Then the system
will track along the lower fixed point until a point when $\gamma$
is greater than $\gamma_L$. At this point, the system will again
jump, this time to a higher fixed-point value. The trademark of
hysterisis is that the two jumps, one when increasing $\gamma$ and
the other when decreasing, occur for different values of $\gamma$.

As is well-known, the full operator region of $\lambda$ phage
contains three sites. We turn briefly to the effect of the
additional site OR1 on the above network. In order to incorporate
its effect, Eq.~(\ref{fast}) must be generalized to account for
additional equilibrium reactions. This generalization amounts to
the incorporation of dimer binding to OR1~\cite{ptashne}, and
permutations of multiple binding possibilities at the three
operator sites. Then, using known relationships between the
cooperative binding rates, the above steps can be repeated and an
equation analogous to Eq.~(\ref{nodimx}) constructed. We obtain
\begin{equation}
\dot x = \frac {\alpha~(2~x^2+50~x^4)}{25+ 29~x^2 + 52~x^4+4~x^6}
- \gamma~x + 1 \label{nodimx2}
\end{equation}
As can be seen, the addition of OR1 has the effect of changing the
first term on the right-hand side of the equation.  While this
augmentation does not affect the qualitative features of the above
discussion, one important quantitative difference is depicted in
Fig.~1B. In this figure, we see that the addition of OR1 has a
large effect on the bistability region, increasing the overall
size of the region by roughly an order of magnitude. Additionally,
the model predicts that, while the drop in the concentration of
repressor at the first bifurcation point will be approximately the
same in both cases, the jump to the higher concentration will be
around five times greater in the system containing OR1. Finally,
since one effect of a larger bistable region is to make the
switching mechanism more robust to noise, these results are of
notable significance in the context of the lysogeny-to-lysis
switching of $\lambda$ phage.

\section*{Additive Noise}
We now focus on parameter values leading to bistability, and
consider how an additive external noise source affects the
production of repressor. Physically, we take the dynamical
variable $x$ described above to represent the repressor
concentration within a colony of cells, and consider the noise to
act on many copies of this colony. In the absence of noise, each
colony will evolve identically to one of the two fixed points, as
discussed above. The presence of a noise source will at times
modify this simple behavior, whereby colony-to-colony fluctuations
can induce novel behavior.

An additive noise source alters the ``background'' repressor
production. As an example, consider the effect of a randomly
varying external field on the biochemical reactions. The field
could, in principle, impact the individual reaction
rates~\cite{xie,astumian}, and since the rate equations are
probabilistic in origin, its influence enters statistically. We
posit that such an effect will be small and can be treated as a
random perturbation to our existing treatment; we envision that
events induced will affect the basal production rate, and that
this will translate to a rapidly varying background repressor
production. In order to introduce this effect, we generalize the
aforementioned model such that random fluctuations enter
Eq.~(\ref{nodimx2}) linearly,
\begin{equation}
\dot{x} = f(x) + \xi(t) \label{langevin}
\end{equation}
where $f(x)$ is the right-hand side of Eq.~(\ref{nodimx2}), and
$\xi(t)$ is a rapidly fluctuating random term with zero mean
($<\xi(t)>=0$). In order to encapsulate the rapid random
fluctuations, we make the standard requirement that the
autocorrelation be ``$\delta$-correlated", i.e., the statistics of
$\xi(t)$ are such that $<\xi(t)\xi(t')>=D\delta(t-t')$, with $D$
proportional to the strength of the perturbation.

Eq.~(\ref{langevin}) can be rewritten as
\begin{equation}
\dot{x} = -\frac{\partial \phi(x)}{\partial x} + \xi(t)
\label{landscape}
\end{equation}
where we introduce the potential $\phi(x)$, which is simply the
integral of the right-hand side of Eq.~(\ref{nodimx}). $\phi(x)$
can be viewed as an ``energy landscape'', whereby $x$ is
considered the position of a particle moving in the landscape. One
such landscape is plotted in Fig.~2A. Note that the stable fixed
values of repressor concentration correspond to the minima of the
potential $\phi$ in Fig.~2A, and the effect of the additive noise
term is to cause random kicks to the particle (system state point)
lying in one of these minima. On occasion, a sequence of kicks may
enable the particle to escape a local minimum and reside in a new
valley.

To solve Eq.~(\ref{landscape}), we introduce the probability
distribution $P(x,t)$, which is effectively the probability of
finding the system in a state with concentration $x$ at time $t$.
Given Eq.~(\ref{landscape}), a Fokker-Planck (FP) equation for
$P(x,t)$ can be constructed \cite{vankampen}
\begin{equation}
\partial_t P(x,t) = -\partial_x (f(x)P(x,t)) +
 \frac{D}{2} {\partial_x}^2 P(x,t) \label{FP}
\end{equation}
We focus here on the value of the steady-state mean (ssm)
concentration. To this end, we first solve for the steady-state
distribution, obtaining
\begin{equation}
P_s(x)=A e^{- \frac{2}{D} \phi(x)}
\end{equation}
where $A$ is a normalization constant determined by requiring the
integral of $P_s(x)$ over all $x$ be unity.  In Fig.~2B, we plot
$P_s(x)$, corresponding to the landscape of Fig.~2A, for two
values of the noise strength $D$.  It can be seen that for the
smaller noise value the probability is distributed around the
lower concentration of repressor, while for the larger noise value
the probability is split and distributed around both
concentrations. This is consistent with our conceptual picture of
the landscape: low noise will enable only transitions from the
upper state to the lower state as random kicks are not sufficient
to climb the steep barrier from the lower state, while high noise
induces transitions between both of the states. Additionally, the
larger noise value leads to a spreading of the distribution, as
expected.

Using the steady-state distribution, the steady-state mean~(ssm)
value of $x \equiv <x>_{ss}$ is given by
\begin{equation}
<x>_{ss} = \int_{0}^{\infty} x A e^{- \frac{2}{D} \phi(x)} dx
\label{uss}
\end{equation}
In Fig.~2C, we plot the ssm concentration as a function of $D$,
obtained by numerically integrating Eq.~(\ref{uss}) and
transforming from the dimensionless variable $x$ to repressor
concentration. It can be seen that the ssm concentration increases
with $D$, corresponding to the increasing likelihood of populating
the upper state, as discussed previously with respect to Figs.~2A
and B.

Figure~2C indicates that the external noise can be used to control
the ssm concentration. As a candidate application, consider the
following protein switch. Given parameter values leading to the
landscape of Fig.~2A, we begin the switch in the ``off'' position
by tuning the noise strength to a very low value.  This will cause
a high population in the lower state, and a correspondingly low
value of the concentration.  Then at some time later, consider
pulsing the system by increasing the noise to some large value for
a short period of time, followed by a decrease back to the
original low value.  The pulse will cause the upper state to
become populated, corresponding to a concentration increase and a
flipping of the switch to the ``on'' position. As the pulse
quickly subsides, the upper state remains populated as the noise
is not of sufficient strength to drive the system across either
barrier~(on relevant time scales). To return the switch to the off
position, the upper-state population needs to be decreased to a
low value. This can be achieved by applying a second noise pulse
of intermediate strength.  This intermediate value is chosen large
enough so as to enhance transitions to the lower state, but small
enough as to remain prohibitive to upper-state transitions.

Figure~2D depicts the time evolution of the switching process for
noise pulses of strengths $D=1.0$ and $D=0.05$. Initially, the
concentration begins at a level of $\mbox{[CI]}=10~\mbox{nM}$,
corresponding to a low noise value of $D=0.01$.  After six hours
in this low state, a $30$-minute noise pulse of strength $D=1.0$
is used to drive the concentration to a value of $\mbox{[CI]} \sim
58~\mbox{nM}$. Following this burst, the noise is returned to its
original value. At $11$ hours, a second $90$-minute pulse of
strength $D=0.05$ is used to return the concentration to its
original value.

\section*{Multiplicative Noise}
We now consider the effect of a noise source which alters the
transcription rate. Although transcription is represented by a
single biochemical reaction, it is actually a complex sequence of
reactions~\cite{vanhippel}, and it is natural to assume that this
part of the gene regulatory sequence is likely to be affected by
fluctuations of many internal or external parameters. We vary the
transcription rate by allowing the parameter $\alpha$ in
Eq.~(\ref{nodimx2}) to vary stochastically, i.e., we set $\alpha
\rightarrow \alpha+\xi(t)$. In this manner, we obtain an equation
describing the evolution of the protein concentration $x$
\begin{equation}
\dot x = h(x) + \xi(t)g(x) \label{multlangevin}
\end{equation}
where $h(x)$ is the right-hand side of Eq.~(\ref{nodimx2}), and
\begin{equation}
g(x)\equiv \frac {2~x^2+50~x^4}{25+ 29~x^2 + 52~x^4+4~x^6}
\end{equation}
Thus, in this case, the noise is multiplicative, as
opposed to additive, as in the previous case.

Qualitatively, we can use the bifurcation plot of Fig.~3A to
anticipate one effect of allowing the parameter $\alpha$ to
fluctuate. Such a bifurcation plot is yet another way of depicting
the behavior seen in Fig.~1A; it can be seen that for certain
values of $\alpha$ there is one unique steady-state value of
repressor concentration, and that for other values there are
three. To incorporate fluctuations, if we envision $\alpha$ to
stochastically vary in the bistable region of Fig.~3A, we notice
that the steep top branch implies the corresponding fluctuations
in repressor concentration will be quite large. This is contrasted
with the flat lower branch, where modest fluctuations in $\alpha$
will induce small variations. In order to verify this observation
quantitatively, we simulated Eq.~(\ref{multlangevin}), the results
of which are presented in Fig.~3B. Beginning with repressor
concentration equal to its upper value of approximately $500$~nM,
we notice that the immediate fluctuations are quite large even
though $\alpha$ varies by only a few percent (Fig.~3A). Then, at
around $700$ minutes, the concentration quickly drops to its lower
value, indicating that the fluctuations envisioned in Fig.~3A were
sufficient to drive the repressor concentration to the dotted line
of Fig.~3A and off the upper branch (across the unstable fixed
point). The final state is then one of very small variation, as
anticipated.

As in the previous section, the steady-state probability
distribution is obtained by transforming Eq.~(\ref{multlangevin})
to an equivalent Fokker-Planck equation \cite{vankampen},
\begin{equation}
\partial_t P(x,t) = -\partial_x (h(x) + \frac{D}{2} g(x)g'(x))P(x,t) +
 \frac{D}{2} {\partial^2_u} g^2(u) P(x,t) \label{FP2}
\end{equation}
where the prime denotes the derivative of $g(x)$ with respect to
$x$. We again solve for the steady-state distribution, obtaining
\begin{equation}
P_s(x)= B e^{-\frac{2}{D} \phi_m(x)} \label{ps2}
\end{equation}
%where $\phi_m(u)$ is given by
%\begin{equation}
%\phi_m(u)=\frac{\delta}{3u^3} -\frac{\gamma}{2u^2} +
%\frac{\alpha+2\beta \delta}{u} -\beta (\alpha+\beta \delta)u +
%\frac{1}{2} \beta^2 \gamma u^2 + 2\beta \gamma Log(u) +
%\frac{D}{2} Log(g(u))
%\end{equation}
As before, the steady-state distribution can be used to obtain the
ssm concentration.

Although not originating from a deterministic equation like that
of Eq.~(\ref{nodimx}), the function $\phi_m(u)$ in Eq.~(\ref{ps2})
can still be viewed as a potential. We now consider parameter
values leading to one such landscape in Fig.~3C. This landscape
implies that we will have two steady-state repressor
concentrations of approximately $5$ and $1200$ nM. This large
difference is due to the largeness of the parameter $\alpha$,
implying that repressor ``induced'' transcription amplifies the
basal rate by a large amount. (Since $d_T$ enters in the numerator
of the definition of $\alpha$, one could construct such a system
experimentally with a high copy-number plasmid). This feature
suggests that multiplicative noise could be used to amplify
protein production, as described in the following example. We
begin with zero protein concentration and very low noise strength
$D$, leading to a highly populated lower state and low overall
concentration. Then, at some later time, we pulse the system by
increasing $D$ for some short interval.  This will cause the upper
state to become quickly populated as it is easy to escape the
shallow valley of the landscape and move into the large basin. In
Fig.~3D, we plot the temporal evolution of the mean repressor
concentration obtained from the simulation of
Eq.~(\ref{multlangevin}). We see that the short noise pulse at
around $20$ hours indeed causes the concentration to increase
abruptly by over three orders of magnitude, making this type of
amplification an interesting case for experimental exploration.

\section*{Discussion}
From an engineering perspective, the control of cellular function
through the design and manipulation of genetic regulatory networks
is an intriguing possibility. In this paper, we have shown how
external noise can be used to control the dynamics of a regulatory
network, and how such control can be practically utilized in the
design of a genetic switch and/or amplifier. Although the main
focus of this work was on a network derived from the promotor
region of $\lambda$ phage, our approach is generally applicable to
any autoregulatory network where a protein-multimer acts as a
transcription factor.

An important element of our control scheme is bistability. This
implies that a necessary criterion in the design of a
noise-controlled applet be that the network is poised in a
bistable region. This could potentially be achieved by methods
such as the utilization of a temperature-dependent repressor
protein, DNA titration, SSRA tagging, or pH control.

Physically, the noise might be generated using an external field.
Importantly, it has been claimed that electromagnetic fields can
exert biological effects~\cite{asbury}. In addition, recent
theoretical \cite{astumian} and experimental \cite{xie} work
suggests a possible mechanism whereby an electric field can alter
an enzyme-catalyzed reaction. These findings suggest that,
although there is global charge neutrality, an external field can
interact with local dipoles which arise through transient
conformational changes or in membrane transport.

Current gene therapy techniques are limited in that transfected
genes are typically either in an ``on" or ``off" state.  However,
for the effective treatment of many diseases, the expression of a
transfected gene needs to be regulated in some systematic fashion.
Thus, the development of externally-controllable noise-based
switches and amplifiers for gene expression could have significant
clinical implications.

~~\\ ~~\\

\noindent ACKNOWLEDGEMENTS. We respectfully acknowledge insightful
discussions with Kurt Wiesenfeld, Farren Issacs, Tim Gardner, and
Peter Jung. This work was supported by the Office of Naval
Research~(Grant N00014-99-1-0554) and the U.S.~Department of
Energy.

\newpage
\clearpage

\section*{Figure Captions}

FIG.~1. Bifurcation plots for the variable $x$ and concentration
of $\lambda$ repressor. (A) Graphical depiction of the fixed
points of Eq.~(\ref{nodimx}), generated by setting $\dot x=0$ and
plotting $\alpha~x^2/(1+2x^2+5x^4)$ and the line $\gamma x -1$. As
the slope $\gamma$ is increased, the system traverses through a
region of multistability and returns to a state of monostability.
(B) Hysterisis loops for the mutant and nonmutant systems obtained
by setting $\dot x=0$ in Eqs.~(\ref{nodimx}) and (\ref{nodimx2}).
Beginning with concentrations of $35$~nM for the mutant system and
$85$ nM for the nonmutant system, we steadily increase the
degradation parameter $\gamma$. In both systems, the concentration
of repressor slowly decreases until a bifurcation point. In the
mutant~(nonmutant) system, the repressor concentration abruptly
drops to a lower value at $\gamma \sim 16$~($\gamma \sim 24$).
Then, upon reversing course and decreasing $\gamma$, the repressor
concentration increases slowly until $\gamma$ encounters a second
bifurcation point at $\gamma \sim 14$~($\gamma \sim 6$), whereby
the concentration immediately jumps to a value of $15$~nM (mutant)
or $70$~nM (nonmutant). The subsequent hysterisis loop is
approximately $10$ times larger in the nonmutant case. Parameter
values are $\alpha=50$, $K_1=0.05~\mbox{nM}^{-1}$, and
$K_2=0.026~\mbox{nM}^{-1}$ for the mutant system, and
$K_2=0.033~\mbox{nM}^{-1}$ for the nonmutant
system~\cite{ptashne}.
\\

\noindent FIG.~2.  Results for additive noise with parameter
values $\alpha=10$ and $\gamma=5.5$. (A) The energy landscape.
Stable equilibrium concentration values of Eq.~(\ref{nodimx2})
correspond to the valleys at $\mbox{[CI]}=10$ and $200$~nM, with
an unstable value at $\mbox{[CI]}=99$~nM. (B) Steady-state
probability distributions for noise strengths of $D=0.04$~(solid
line) and $D=0.4$~(dotted line). (C) The steady-state equilibrium
protein concentration plotted versus noise strength. The
concentration increases as the noise causes the upper state of (A)
to become increasingly populated. (D) Simulation of
Eq.~(\ref{langevin}) demonstrating the utilization of external
noise for protein switching. Initially, the concentration begins
at a level of $\mbox{[CI]}=10$~nM corresponding to a low noise
value of $D=0.01$. After six hours, a large $30$-minute noise
pulse of strength $D=1.0$ is used to drive the concentration to
$58$~nM. Following this pulse, the noise is returned to its
original value. At $11$ hours, a smaller $90$-minute noise pulse
of strength $D=0.04$ is used to return the concentration to near
its original value. The simulation technique is that of
Ref.~\cite{sancho}.
\\

\noindent FIG.~3.  Results for multiplicative noise. (A)
Bifurcation plot for the repressor concentration versus the model
parameter $\alpha$. The steep upper branch implies that modest
fluctuations in $\alpha$ will cause large fluctuations around the
upper fixed value of repressor concentration, while the flat lower
branch implies small fluctuations about the lower value. (B) The
evolution of the repressor concentration in a single colony,
obtained by simulation of Eq.~(\ref{multlangevin}). Relatively
small random variations of the parameter $\alpha$ ($\sim$ 6\%)
induce large fluctuations in the steady-state concentration until
around $700$ minutes and small fluctuations thereafter. (C) Energy
landscape for parameter values $\alpha=100$ and $\gamma=8.5$. (D)
Large-scale amplification of the protein concentration obtained by
simulation of Eq.~(\ref{multlangevin}). At $20$ hours, a
$60$-minute noise pulse of strength $D=1.0$ is used to quickly
increase the protein concentration by over three orders of
magnitude. The parameter values are the same as those in (C).

%\end{document}

\newpage

\clearpage

\pagestyle{empty}

\begin{figure}
\center

\epsfig{file=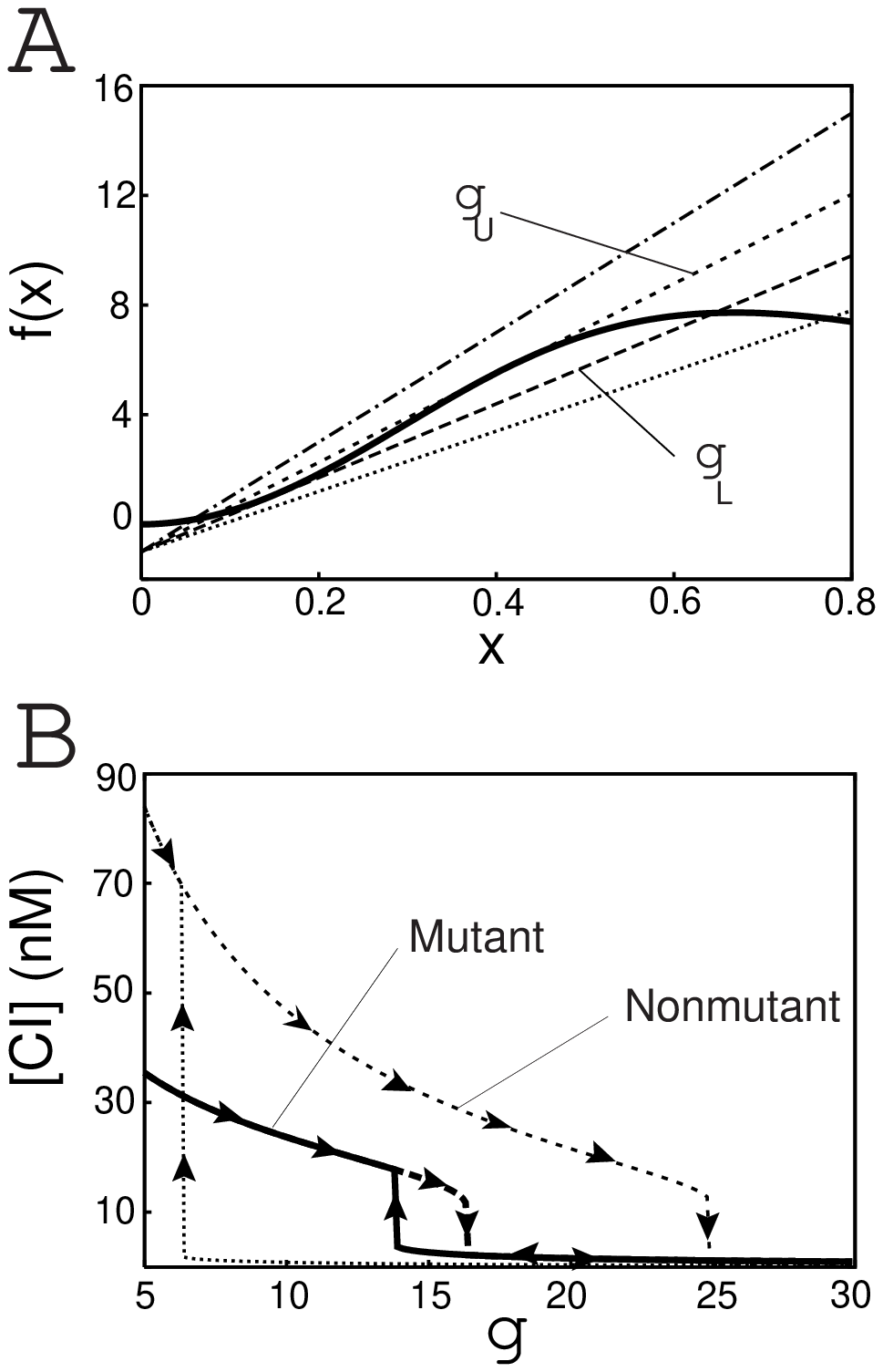,height=4in}
\end{figure}
\begin{center}
~\\~\\~\\~\\~\\~\\~\\~\\Figure 1 - Hasty et al.
\end{center}

\newpage
\clearpage

\pagestyle{empty}
\begin{figure}
\center

\epsfig{file=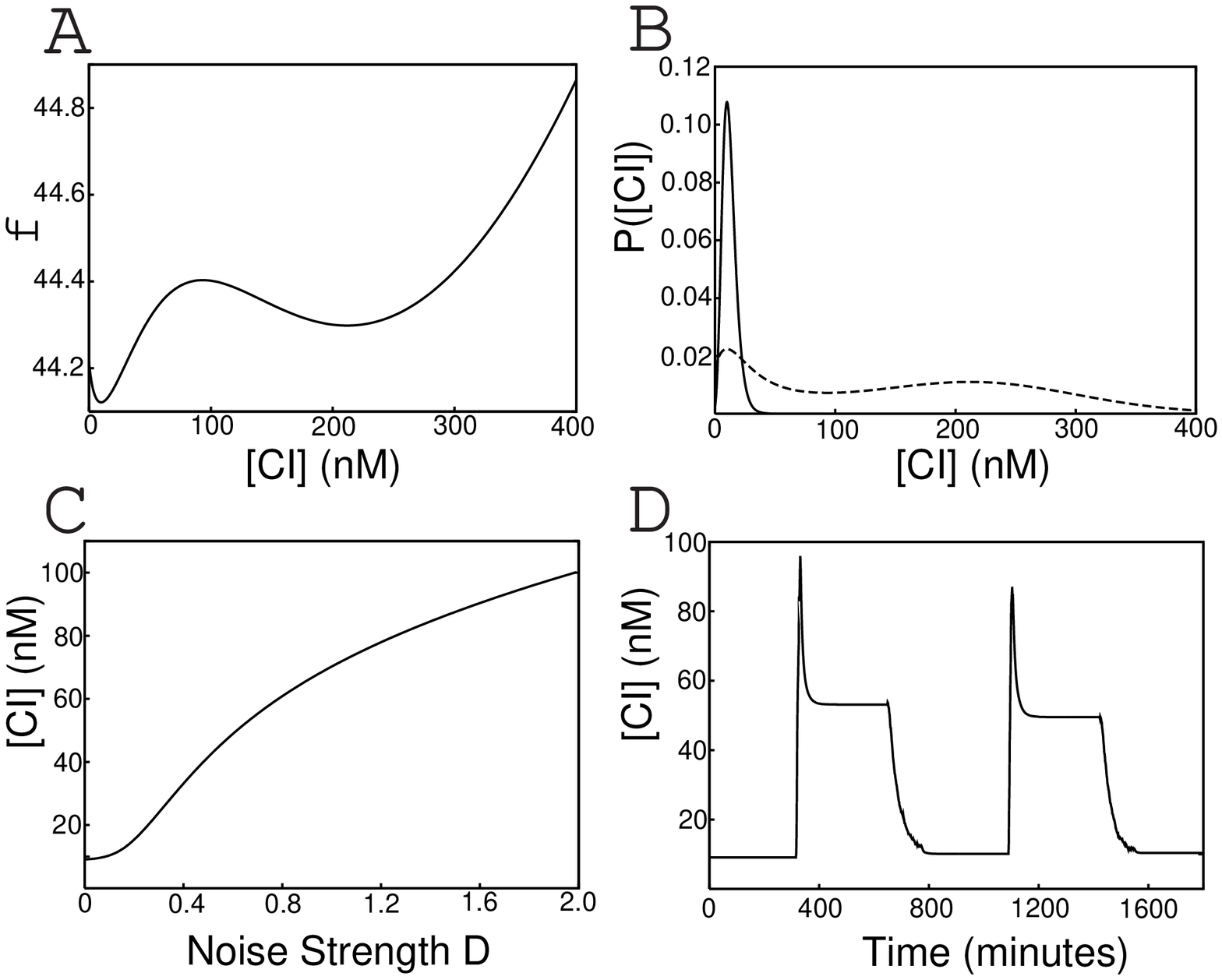,height=4in}
\end{figure}

\begin{center}
~\\~\\~\\~\\~\\~\\~\\~\\Figure 2 - Hasty et al.
\end{center}

\newpage
\clearpage
\pagestyle{empty}
\begin{figure}
\center

\epsfig{file=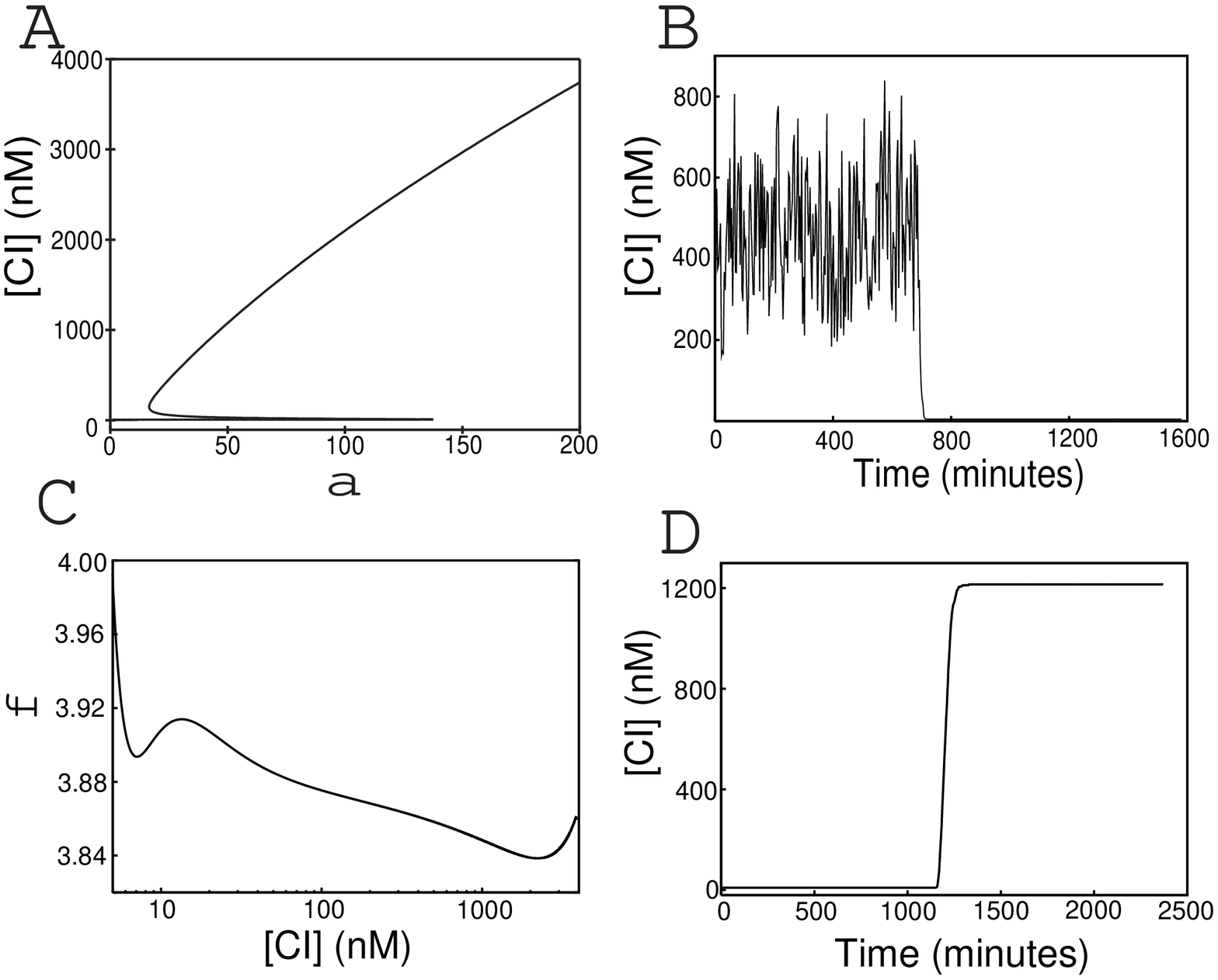,height=4in}
\end{figure}

\begin{center}
~\\~\\~\\~\\~\\~\\~\\~\\Figure 3 - Hasty et al.
\end{center}


\begin{thebibliography}{xx}
\bibitem{Dickson} Dickson,~R., Abelson,~J., \& Barnes,~W.~(1975)
{\em Science} {\bf 187}, 27--35.

\bibitem{Davidson1} Yuh,~C.~H., Bolouri,~H., \&
Davidson,~E.~H.~(1998) {\em Science} {\bf 279}, 1896--1902.

\bibitem{GenesVI} Lewin,~B.~(1997) in {\em Genes VI} (Oxford University Press,
Oxford).

%\bibitem{Hanawalt} Hanawalt,~P.~C.~(1998)
%{\em Mut. Res.} {\bf 400}, 117--125.

\bibitem{harada} Harada,~Y. et. al.~(1999) {\em Biophys. J.} {\bf
76}, 709--715.

\bibitem{arkin} McAdams,~H.~H. \& Arkin,~A.~(1997)
{\em Proc. Natl. Acad. Sci.} {\bf 94}, 814--819.

\bibitem{shapiro} McAdams,~H.~H. \& Shapiro,~L.~(1995)
{\em Science} {\bf 269}, 650--656.

\bibitem{keller} Keller,~A.~(1995) {\em J. Theor. Biol.} {\bf
172}, 169--185; Smolen,~P, Baxter,~D.~A., \& Byrne,~J.~H.~(1998)
{\em Am. J. Physiol.--Cell Ph.} {\bf 43}, C531--C542; Wolf,~D.~M.
\& Eeckman,~F.~H.~(1998) {\em J. Theor. Biol.} {\bf 195},
167--186.

\bibitem{horsthemke} Horsthemke,~W. \& Lefever,~R.~(1984) in
{\em Noise-Induced Transitions} (Springer-Verlag, Berlin).

\bibitem{collins} Gardner,~T.~S., Cantor,~C.~R., \& Collins,~J.~J.~(1999)
{\em Nature}, in press.

\bibitem{ptashne} Ptashne,~M et al.~(1980) {\em Cell} {\bf 19}, 1--11;
Johnson,~A.~D. et al.~(1981) {\em Nature} {\bf 294}, 217--223.

\bibitem{johnson} Johnson,~A.~D. et al.~(1980) {\em Methods Enzymol.} {\bf 65},
839--856.

\bibitem{xie} Xie,~T.~D., Marszalek,~P., \& Chen,~Y.~(1994)
{\em Biophys. J.} {\bf 67}, 1247--1251.

\bibitem{astumian} Astumian,~R.~D. \& Robertson,~B.~(1993)
{\em J. Am. Chem. Soc.} {\bf 115}, 11063--11068.

\bibitem{vankampen} Van Kampen,~N.~G.~(1992) in {\em Stochastic Processes in
              Physics and Chemistry} (North-Holland, Amsterdam).



\bibitem{asbury} See, for example, Berg,~H.~(1995) {\em
Bioelectrochem. Bioenerg.} {\bf 38}, 153--159; Liu,~D.~S. et
al.~(1990) {\em J. Biol. Chem.} {\bf 265}, 7260--7271;
Otter,~M.~W., McLeod,~K.~J., \& Rubin,~C.T.~(1998) {\em Clin.
Orthop.} {\bf 355}, S90--S104; Asbury,~C.~L. \& van den
Engh,~G.~(1998) {\em Biophys. J.} {\bf 74}, 1024--1130.



\bibitem{vanhippel} von Hippel,~P.~H.~(1998) {\em Science} {\bf 281}, 660--665.

\bibitem{sancho} Sancho,~J., Miguel,~M.~S., \& Katz,~S.~(1982) {\em Phys. Rev. A} {\bf 26}, 1589--1593.

\end{thebibliography}
\end{document}